\magnification=\magstep1

\def\bl{\bigskip}
\centerline{{\bf Asymmetric neutrino emision and formation of rapidly
moving pulsars}}
\bl
\centerline{{\bf G.S.Bisnovatyi-Kogan}}
\bl
\bl
Head title: Asymmetric neutrino emission.
\bl
\noindent
{\bf Abstract}
\bl
\noindent
The neutron star formation during the collapse with
the strong magnetic field
may lead to a mirror symmetry violation and formation of an asymmetric
magnetic field. Dependence of the week interaction cross-section on the
magnetic field strength lead to the asymmetric neutrino flux and formation
of rapidly mooving pulsars due to the recoil action as well as rapidly
moving black holes.
\bl
\noindent
{\bf 1. Introduction}
\bl
\noindent
Existence of the pulsars moving with the velocities up to 500 km/s
(Harrison, Lyne and Anderson, 1991) is a big
challenge to the theory of the neutron star formation in the spherically
symmetrical collapse. Collapse of the rotating star with
an axial symmetry make no changes and the
consideration of the Blaauw effect during the formation of pulsars
in the binaries cannot produce such a high speeds. The most plausible
explanation for the birth of rapidly mooving pulsars seems to be a
suggestion of the kick at the birth from the asymmetric explosion
(Shklovskii, 1970, see also Radhakrishnan, 1991). Here the estimations are
made for the strength of the kick, produced by the asymmetric neutrino
emission during the collapse. Chugay (1984) and Dorofeev, Rodionov and Ternov
(1985) tried to produce the asymmetric neutrino pulse due to the action of
the strong magnetic field, which make the electrons to be polarized. It
leads to the mirror asymmetric neutrino emission due to the P symmetry violation
in the week interaction Lagrangian. As was shown by Bisnovatyi-Kogan
(1989), the strength of the poloidal magnetic field needed for the explanation
of the visible velocities is much larger, then the fields, observed in
theese pulsars.
\hfill\break\indent
The asymmetry of the neutrino pulse, considered here is produced by the
asymmetry of the field distribution, formed during the collapse and
differential rotation (Bisnovatyi-Kogan and Moiseenko, 1992). Qualitative
picture of the neutrino pulse asymmetry due to the dependence of the neutrino
cross-section on the magnetic field strength was considered by
Bisnovatyi-Kogan (1991). Here the quantitative estimations are presented.
\bl
\noindent
{\bf 2. Formation of the asymmetric magnetic field distribution}
\bl
\noindent
In order to obtain the mirror asymmetric magnetic field distribution let us
consider a rotating presupernova star with the dipole poloidal and symmetric
toroidal field. Collapse of such star after the loss of stability lead to the
formation of the rapidly and differentially rotating neutron star. The field
amplification during the differential rotation lead to
the formation of the additional
toroidal field from the poloidal one. The toroidal field, made from the
dipole poloidal by twisting of the field lines is antisymmetric
relative to the symmetry plane. The sum of the initial symmetric with
the induced antisymmetric toroidal fields has no plane symmetry
and the field in one hemisphere is larger then in another one.
Such symmetry violation happens always when the star begins to rotate
differentially and posesses toroidal and poloidal fields with the
different symmetry properties (dipole poloidal and symmetric toroidal; or
quadrupole poloidal with antisymmetric toroidal; etc).
\hfill\break\indent
In the absence of the
dissipative processes with the perfect magnetic field freezing
the neutron star returns to the state of rigid rotation loosing the
induced toroidal field and restoring mirror symmetry of the matter
distribution. In the presence of the field dissipation
the rigidly rotating star returns to the rigid rotation
having an asymmetric toroidal field and an asymmetry of the
matter distribution. The formation of
the asymmetric toroidal field distribution
may be followed by the asymmetric magnetorotational explosion, producing
the neutron star recoil and the rapidly mooving star (Bisnovatyi-Kogan, 1970;
Ardelyan et. al.,1979; Bisnovatyi-Kogan, Moiseenko,1992).
Even in the case, when the magnetorotational explosion is not
effective, the neutron star accelleration may happen due to the dependence
of the cross-section of the week interactions on the magnetic field.
\hfill\break\indent
Influence of the magnetic field on the processes of week interaction was studied
well  for the process of the
neutron decay (O'Connel, Matese,1969). The influence
begins, when the characteristic energy of the electron on the
Landau level with the
Larmor rotation ${\hbar eB \over m_e c}$ becomes of the order of the energy
of the decay, which for the neutron decay is of the order of $m_e c^2$. The
equality of these energies determines the critical magnetic field
$$ B_c={m_e^2 c^3 \over e \hbar} = 4.4 \times 10^{13} {\rm Gs} \eqno(1) $$
The probability of the neutron decay $ W_n $
in the strong magnetic field without the matter is
$$ W_n=W_0[1+0.17(B/B_c)^2+...] \quad {\rm at} \quad B \ll B_c $$
$$ W_n=0.77 W_0 (B/B_c) \quad {\rm at } \quad B \gg B_c    \eqno (2) $$
\noindent
The formulas (2) may be easily
generalized for the neutron decay and the electron capture
in the presence of matter with Fermi distribution, what needs only
changes of the phase volume of the integration. For the fully
degenerate electrons the integration over the phase volume may be
done analitically (see, e.g. Shulman,1977). In the
fully degenerate case  the smooth
dependence of the decay or capture probabilities on the field strength
is accompanied by the jumps of the derivatives
when the difference between the energy of the decay and the Fermi energy of
the electrones connected with the motion along the maghetic field
crosses the energy of the corresponding Landau level.
The neutrino emissivity from the synchrotron  and $ e^+e^- $ annihilation
processes in the nonrelativistic limit were studied by Kaminker et.al.
(1991).
\hfill\break\indent
After the collapse of the rapidly rotating star the new forming neutron star
rotates with the period $P$ about 1 ms, corresponding to the critical
rotational velocity. Differential rotation leads to the linear
amplification of the toroidal field, according to the approximate
law
$$ B_{\phi}=B_{\phi 0}+B_p (t/P)   \eqno (3) $$
Numerical calculations of the spherically symmetrical collapse gave
(Nadjozhin, 1978) several tens of seconds for the
time of the neutrino emission. This time may even increase with an account of
the rotation. During 20 s the induced toroidal magnetic field
will become equal to $ 2 \times 10^4 B_p $, corresponding to $ 10^{15}
\div 10^{17} $ Gs for $ B_p = 10^{11} \div 10^{13} $ Gs, observed in the
pulsars. Adopting the initial toroidal field equal
to $B_{\phi 0}=(10 \div 10^3) B_p = 10^{12} \div 10^{16} $, we may start the
estimation of the asymmetry of the neutrino pulse produced by the anisotropic
neutrino emission. It is easy to see, that for symmetric $ B_{\phi,0} $
and dipole poloidal field the difference
$\Delta B_{\phi} $ between the magnetic
fields absolute values
in two hemispheres increases, until it reaches the value
$ 2 B_{\phi 0} $. It remains constant later, while the relative
difference
$$\delta_B={\Delta B_{\phi} \over B_{\phi +} + B_{\phi -} }$$
\noindent
decreases.
\bl
\noindent
{\bf 3. Neutrino heat conductivity and energy losses }
\bl
\noindent
The main neutrino flux is formed in the region where the mean free path of
the neutrino is less then the stellar radius. The neutrino energy flux,
$ H_{\nu} $ connected
with the temperature gradient may be written as (Imshennik, Nadjozhin, 1972)
$$ H_{\nu}=-{7 \over 8} {4acT^3 \over 3} l_T {\partial T \over \partial r}
 \eqno (4) $$
\noindent
Here was neglected the part of the
heat flux, connected with the gradient of the lepton
charge. In order to estimate the neutrino flux distribution over the surface
of the star we consider for simplicity a set of spherically symmetrical
stars with different neutrino opacity distributions and the same central
temperature. The value $l_T$ having a sence of the neutrino mean free path
is connected with the neutrino opacity
$ \kappa_{\nu} $ as
$$\kappa_{\nu}=1/(l_T \rho) \eqno (5) $$
\noindent
Calculations of the spherically symmetrical collapse (Nadjozhin, 1978) have
shown, that during the phase of the main neutrino emission the hot neutron star
consists
of the quasiuniform quasiisothermal core with the temperature $T_i$,
the mass of which increases with
the time, and the region between the neutrinosphere and
the isothermal core, where the
temperature smoothly decreases in about 10 times and
behaviour of density, which finally drops about 6 times is nonmonotoneous.
In this
region, containing about one half of the neutron star mass, neutrino
flux is forming. We suggest for sumplicity in this region the power
dependences for the temperature
$$ T=T_i \left( {r_i \over r} \right)^m  \eqno (6) $$
\noindent
and for $ l_T $
$$ l_T={1 \over \kappa \rho }=l_{Ti} \left(
{r \over r_i} \right)^n \eqno (7) $$
\noindent
The neutrinosphere with the radius $r_{\nu} $
is determined approximately by the relation
$$ \int_{r_{\nu}}^{\infty} \kappa_{\nu} \rho dr = \int_
{r_{\nu}}^{\infty} {dr \over l_T}=1    \eqno (8) $$
\noindent
Using the distribution (7) outside the neutrinosphere we get from (8) the
relation
$$ r_{\nu}=r_i{ \left( r_i \over (n-1)l_{Ti}
\right) }^{1 \over n-1} \eqno (9) $$
\noindent
From (4)-(7), using (9) we get the temperature of the neutrinosphere $T_{\nu}$
and the heat flux on this level $H_{\nu}$, which outside the neutrinosphere
is approximately $\sim r^{-2}$, corresponding to the constant neutrino
luminosity $L_{\nu}$
$$ T_{\nu}=T_i \left( (n-1)l_{Ti} \over r_i \right)^{m \over n-1} \eqno (10)$$
$$ L_{\nu}=4\pi r_{\nu}^2 H_{\nu}={7 \over 8} m {16 \pi acT_i^4 \over 3}
(n-1)^{4m-n+1 \over n-1} r_i^2 \left(l_{Ti} \over r_i \right)
^{4m-2 \over n-1}    \eqno (11) $$
\noindent
In order to estimate the anisotropy of the neutrino flux we compare
two stars with the same radius and temperature of the core $ r_i$
and $ T_i$ and different opacities (different $l_{Ti}$ ). Consider
for simplicity a star where $l_{Ti}$ is different
and constant in two hemispheres, the laws (6) and (7) are the same
and each hemisphere radiate independently with the luminosities, equal to
one half of (11) with different $l_{Ti}$. The anisotropy of the flux
$$ \delta_L ={L_+-L_- \over L_++L_-} \eqno (12) $$
\noindent
with $L_+$ and $L_-$ as luminosities in the different hemispheres, may be
calculated, using (11). For small difference between hemispheres we
get from (11)
$$ \delta_L={\Delta L \over L}={4m-2 \over n-1} {\Delta l_{Ti} \over
l_{Ti}}   \eqno (13) $$
\noindent
It is clear from (10), that neutrinosphere exist only at $n > 1$.
It follows from (11), that when $m={1 \over 2} $ the neutrino fluxes in both
hemispheres are equal because smaller opacity and larger neutrinosphere
temperature $T_{\nu}$ from (10) is compensated by smaller neutrinosphere
radius $r_{\nu}$ from (9), so that the luminosity, determined by the
product $ T_{\nu}^4 r_{\nu}^2 \sim T_{\nu}^4 r_{\nu} l_{T\nu} $
is constant. For $m>{1 \over 2} $
the larger $ l_{Ti} $ corresponds to the larger
luminosity, what means the total excess of the more energetic neutrino,
and the opposite situation happens for $m<{1 \over 2} $. Let us emphasize
that this conclusion is valid only for the same power dependences
(6) and (7) for $T$ and $l_T$ with different values of $l_{Ti}$ at
the boundary of the isothermal core. It is not possible to apply this
conclusion directly for the different opacity laws, like in the case
of the stellar neutrino luminosity in different neutrino sorts (electron,
muon and tau).
\bl
\noindent
{\bf 4. Neutron star acceleration by anisotropic neutrino pulse.}
\bl
\noindent
The equation of motion of the neutron star with the
mass $ M_n $ radiating the anisotropic neutrino flux is written as
$$ M_n {dv_n \over dt} = {1 \over c } \int_0^{\pi} L_{\nu}(t,\theta)
\cos \theta d \theta \eqno (14) $$
\noindent
The total neutrino luminosity
$$ L_{\nu}(t)=\int_0^{\pi} L_{\nu}(t,\theta) d \theta \eqno (15) $$
\noindent
may be taken from the spherically symmetrical calculations of Nadjozhin (1978)
or Mayle et.al (1987). Consider for simplicity, that the neutrino fluxes in
upper $L_{+}$ and lower $L_{-}$ hemispheres are constant over $\theta$.
Then (14),(15) may be written as
$$M_n{dv_n \over dt }= {L_{+}-L_{-} \over c}  \eqno (16) $$
$$L_{+}+L_{-}= {2 \over \pi} L_{\nu}(t)  \eqno (17) $$
\noindent
For the power distributions (9),(10) it follows from (11) the dependences
$$ L_{\pm}=A l_{Ti\pm}^{4m-2 \over n-1}  \eqno (18) $$
\noindent
where $l_{Ti\pm}$ are the average values of $I_{Ti}$ in two hemispheres.
In general $l_{Ti}$ is determined by different neutrino processes and
depends on $B$.
\hfill\break\indent
As an example consider the dependence on $B$ in the form (2). Making
interpolation between two asimptotics we get dependence
$$ l_{Ti\pm} \sim {1 \over W} =l_{T0} {1+{ \left (B \over B_c \right )}^3
\over 1+0.17 {\left( B \over B_c \right) }^2+0.77{\left( B \over B_c
\right) }^4}  \eqno (19) $$
\noindent
The time dependence of the average value of $B$ in each hemisphere may be
found from (3), taking average values of $B_{\phi 0 \pm}$
and average values of $B_{p \pm} $, so that
$$ B_{p+}=-B_{p-}, \quad  B_{\phi 0 +}=B{\phi 0 -} \eqno (20)$$
\noindent
The values of $l_{Ti \pm}$ in each hemisphere may be written as
$$ L_{\pm}=A l_{T0}^{4m-2 \over n-1} F_{\pm}= D(t) F_{\pm} \eqno (21) $$
\noindent
where $F{\pm}$ as a function of $B$ may be found from the comparison with (18),
(19) and $B$ as a function of time is taken from (3) with account of (20).
From (17),(21) we get
$$ D(t)={2 L_{\nu}(t) \over \pi (F_++F_-)} \eqno (22) $$
\noindent
The equation (16) with account of (21),(22) may be finaly written as
$$ M_n {dv_n \over dt}={2 \over \pi} {L_{\nu} \over c} {F_--F_+ \over
F_-+F_+} \eqno (23) $$
\noindent
where the time dependence of $L_{\nu}$ is taken from the spherically symmetrical
calculations of the collapse and nondimentional time functions $F{\pm}$
are determined by the structure of the neutron star above the isothermal core
and the average time dependence of $B_{\phi}$ in two hemispheres.
\bl
\noindent
{\bf 5. Numerical estimations }
\bl
\noindent
Consider for simplicity the distributions (6),(7) with
$${4m-2 \over n-1}=1 \eqno (24) $$
\noindent
The main acceleration of the neutron star occures when $B \gg B_c$,
so the functions $F_{\pm}$ reduce to
$$ F_{\pm}={B_c \over 0.77 B_{\pm} } \eqno (25) $$
\noindent
and the equation of motion (23) may be written as
$$ M_n {dv_n \over dt}={2 \over \pi} {L_{\nu} \over c} {\vert B_+\vert -
\vert B_-\vert \over \vert B_+\vert +\vert B_-\vert } \eqno (26) $$
\noindent
For linear functions
$$ B_{\pm} \equiv B_{\phi \pm}=a \pm bt  \eqno (27) $$
$$ a=B_{\phi 0}, b={\vert B_p \vert \over P} $$
\noindent
with $a$ and $b$ determining by (3),(20). Take constant $ L_{\nu} $
$$ L_{\nu}={0.1 M_n c^2 \over 20 s} \eqno (28) $$
\noindent
With these simplifications the final velocity of the neutron star $v_{nf}$
as a result of the solution of (26) is written in the form
$$ v_{nf}={2 \over \pi} {L_{\nu} \over M_n c}{PB_{\phi 0} \over \vert B_p
\vert} (0.5+\ln \left( {20 \, s
\over P} {\vert B_p \vert \over B_{\phi 0}} \right) )\eqno (29)$$
\noindent
For $P=10^{-3} \, s $ and $L_{\nu}$ from (28) we obtain from (29)
$$ v_{nf}={2 \over \pi} {c \over 10} {P \over 20 \, s} x(0.5+
\ln({20 \, s \over P} {1 \over x} )) \approx 1 {km \over s} x(0.5+\ln
{2\times 10^4 \over x}) \eqno (30) $$
\noindent
For the value $ x={B_{\phi 0} \over \vert B_p \vert} $ between $20$
and $10^3$, we have $v_{nf}$ between 140 and 3000 km/s, what can explain
the most rapidly mooving pulsars. The formula (30) may be applied when
$ B_{\phi 0} \gg B_c $ and $ x \gg 1$.
\hfill\break\indent
For nonlinear dependence in (18) the analitical estimation of $v_n$ may
be done, when the main acceleration happens when
$$bt \gg a \eqno (30) $$
\noindent
in (27).
In the same conditions $B_{\pm} \gg B_c $ we have
$$ F_{\pm}={\left(B_c\over 0.77 B_{\pm}\right)}^{4m-2 \over n-1} \eqno (31) $$
\noindent
Making expansion in (31), using (27),(30), we obtain from (23)
$$ M_n {dv_n \over dt}={2 \over \pi} {L_{\nu} \over c}
{4m-2 \over n-1} x {P \over t} \eqno (32) $$
\noindent
Integration of (32) gives the result which differs from (30)
approxomately only by multiplicator $(4m-2)/(n-1)$
$$ v_{nf}={2 \over \pi} {c \over 10} {P \over 20 \, s} x {4m-2 \over
n-1} \ln({20 \, s \over P} {1 \over x} )
\approx 1 {km \over s} {4m-2 \over n-1}
x \ln {2\times 10^4 \over x} \eqno (33) $$
Acceleration of the collapsing star by anisotripic neutrino emission may
be done even when the star does not stop on the stage of the neutron star
and collapses to the black hole. We may expect black holes of stellar
origin moving with rapid velocities, like radiopulsars. It means, that
they may be situated much higher over the galactic disk, then their
progenitor - very massive stars.
\bl
\noindent
{\bf 6. Conclusion.}
\bl
\noindent
The results, obtained above have shown, that the anisotropy of the neutrino
pulse, produced by the mirror asymmetric magnetic field distribution may
explain the observed high velocities of radiopulsars for
realistic magnetic field strengthes. This mechanism
of acceleration acts in all cases of anisotropic neutrino emission,
including the case of the formation of black hole. It means, that black
holes could be found on the distance from the galactic plane much
larger, then the thickness of the Galactic disc, containing very massive
stars.
\bl
\noindent
{\bf References}
\bl
\noindent
Ardelyan N.V., Bisnovatyi-Kogan G.S., Popov Yu.P., 1979, Astron.Zh.,
{\bf 56}, 1244.
\hfill\break
Bisnovatyi-Kogan G.S., 1970, Astron. Zh., {\bf 47}, 813.
\hfill\break
Bisnovatyi-Kogan G.S., 1989, "Fizicheskie voprosy teorii zvezdnoj evolyuzii",
\hfill\break\indent
Nauka, Moskva (in Russian).
\hfill\break
Bisnovatyi-Kogan G.S., 1991, in Proc. 6-th Capri Workshop "Stellar Jets and
\hfill\break\indent
bipolar outflows" (in press).
\hfill\break
Bisnovatyi-Kogan G.S.,Moiseenko S.G., 1992, Astron Zh. (in press).
\hfill\break
Chugay N.N., 1984, Pisma Astron Zh., {\bf 10}, 210.
\hfill\break
Dorofeev O.F., Rodionov V.N., Ternov I.M., 1985, Pisma Astron.Zh, {\bf
11}, 302.
\hfill\break
Harrison P.A., Lyne A.G., Anderson B., 1991, in Proc. NATO-ARW  "X-ray
\hfill\break\indent
Binaries and Formation of Binary and Millisecond Pulsars", Kluwer (in press).
\hfill\break
Imshennik V.S., Nadjozhin D.K., 1972, Zh. Exper. Theor. Phys., {\bf 63}, 1548.
\hfill\break
Kaminker A.D., Levenfish K.R., Yakovlev D.G., Amsterdamski P.,Haensel P.,
\hfill\break\indent
1991, CAMK preprint 233, July.
\hfill\break
Mayle R., Wilson J.R., Schramm D.N., 1987, Astrophys.J., {\bf 318}, 288.
\hfill\break
Nadjozhin D.K.,1978, Astrophys. Space Sci., {\bf 53}, 131.
\hfill\break
O'Connel R., Matese J., 1969, Nature, {\bf 222}, 649.
\hfill\break
Radhakrishnan S., 1991, in Proc. NATO-ARW "X-ray Binaries and Formation
\hfill\break\indent
of Binary and Millisecond Pulsars", Kluwer (in press).
\hfill\break
Shklovskii I.S., 1970, Astron.Zh., {\bf 46}, 715.
\hfill\break
Shulman G.A., 1977, Astrofizika, {\bf 13}, 657.

\end